\documentclass[aps,prl,twocolumn,superscriptaddress,noeprint,longbibliography,floatfix]{revtex4-2}

\usepackage{graphicx}
\usepackage{bm}
\usepackage{amsmath, amssymb}
\usepackage{booktabs}
\usepackage{mathrsfs}
\usepackage[normalem]{ulem}
\usepackage{pifont}
\usepackage[mathscr]{euscript}
\usepackage{hyperref}
\usepackage{soul}
\usepackage{dsfont}
\usepackage[inline]{enumitem} 
\usepackage[dvipsnames]{xcolor}
\usepackage{tikz}

\definecolor{DarkerGray}{HTML}{a1a1a1}
\definecolor{LiteGray}{HTML}{e7e7e7}

\tikzset{matter spin/.style={circle,thick,draw,DarkerGray,line width=1,fill=LiteGray,scale=0.75}}

\hypersetup{
    colorlinks = true,
    linkcolor  = blue,
    citecolor  = blue,
    urlcolor   = blue,
    linktocpage=true
}

\usepackage{physics}
\usepackage{comment}

\newcommand{\add}[1]{ { \color{blue}  #1 }}

\begin{document}

\title{Probing anyonic statistics via Mach-Zehnder interferometry in quantum computers}

\author{Shiyu Zhou}
\affiliation{Department of Physics, Boston University, Boston, MA, 02215, USA}
\affiliation{Perimeter Institute for Theoretical Physics, Waterloo, Ontario, Canada N2L 2Y5}

\author{Yi Teng}
\affiliation{TCM Group, Cavendish Laboratory, University of Cambridge, Cambridge CB3 0HE, UK}

\author{Claudio Chamon}
\affiliation{Department of Physics, Boston University, Boston, MA, 02215, USA}

\author{Claudio Castelnovo}
\address{TCM Group, Cavendish Laboratory, University of Cambridge, Cambridge CB3 0HE, UK}

\author{Armin Rahmani}
\affiliation{Department of Physics and Astronomy and Advanced Materials Science and Engineering Center, Western Washington University, Bellingham, Washington 98225, USA}


\begin{abstract}
We introduce a synthetic Mach-Zehnder interferometer for digitized quantum computing devices to probe fractional exchange statistics of anyonic excitations that appear in quantum spin liquids. Employing an IonQ quantum computer, we apply this scheme to the toric ladder, a quasi-one-dimensional version of the toric code. We observe interference patterns resulting from the movement of `electric' excitations in the presence and absence of `magnetic' ones. We model the noise in IonQ via depolarizing Lindbladian dynamics, and find quantitative agreement with the measurements obtained from the quantum device. The synthetic Mach-Zehnder interferometer can thus also serve as an effective means to probe the coherence length and time scales of multi-qubit noisy quantum devices.
\end{abstract}

\maketitle
%
%

Interferometry has served as a powerful experimental technique, leading to numerous significant discoveries in physics. The Michelson interferometer was central to the Michelson-Morley experiment~\cite{Michelson_Morley1887}, which revealed that light propagates at constant speed in perpendicular directions, contradicting the existence of an ether through which light was thought to travel. Combinations of Michelson and Fabry-Pérot interferometers were instrumental in detecting gravitational waves~\cite{detection_gravitational_waves}. In particle systems, recent experiments employing Fabry-Pérot interferometers have shown fractional exchange statistics in quasiparticles of certain quantum Hall states~\cite{Nakamura2020,Nakamura2022,Nakamura2023}. Other interferometric setups, such as Mach-Zehnder interferometers, have been theoretically studied~\cite{Law2006, Batra_2023} and experimentally implemented~\cite{Kundu2023} to explore fractional exchange statistics in quantum Hall states. Interferometry of thermal currents of anyons has also been proposed to probe fractional statistics in quantum spin liquids~\cite{PhysRevLett.127.167204,PhysRevB.107.104406}.

In this paper, we design a synthetic Mach-Zehnder interferometer to demonstrate fractional exchange statistics of excitations in topological states of matter~\cite{moore_moessner} realizable in quantum computers and simulators~\cite{Kirmani_2022, Kirmani_2023, Lopez_Bezanilla_2023}. Examples of these topological systems include quantum spin liquids --- a central topic in quantum many-body physics that is challenging to investigate in material systems~\cite{Knolle_2019, Broholm_2020}. There has been much effort and success in implementing such states in existing quantum devices~\cite{Satzinger2021, Semeghini_2021, iqbal2023topological, Iqbal_2024}. Here, we focus on simulating the nonequilibrium dynamics, generated by a \textit{time-independent} Hamiltonian, which leads to the propagation of fractionalized excitations characteristic of quantum spin liquids and the associated interferometric signatures. This approach contrasts with the more conventional method of producing approximate quantum eigenstates and adiabatically manipulating excitations.

Our setup has the following features: (1) the qubits in the quantum device correspond directly to the physical degrees of freedom of the model that hosts the anyonic excitations; and (2) the quantum circuit performs a Trotterized representation of unitary evolution generated by the time-independent Hamiltonian of the same model. This results in natural spinon propagation, eliminating the need to impart any initial momentum to the excitations or guide the spinons along specific paths. The significance of our method rests in its ability to leverage synthetic quantum platforms for a faithful emulation and identification of the distinctive signatures of anyonic statistics during nonequilibrium quantum evolution. Artificial Mach-Zehnder interferometry can effectively assess coherence lengths and timescales in quantum computing platforms, offering insights into their intrinsic noise characteristics.

{
To demonstrate our approach, we investigate a quasi-1D ladder variant of the 2D toric code~\cite{Kitaev2003}, which supports two types of excitations: spinons and visons, representing `electric' and `magnetic' charges, respectively, which exhibit mutual semionic statistics. We implement this setup on the IonQ quantum device and successfully observe signatures of mutual semionic statistics in the propagation of spinons, both in the presence and absence of a stationary background of visons. This quasi-1D ladder is a carefully adapted version of the 2D toric code, designed to enhance the observable effects of fractional statistics in available quantum devices.
%
%

%
\begin{figure*}[!t]
    \includegraphics[width=0.9\textwidth]{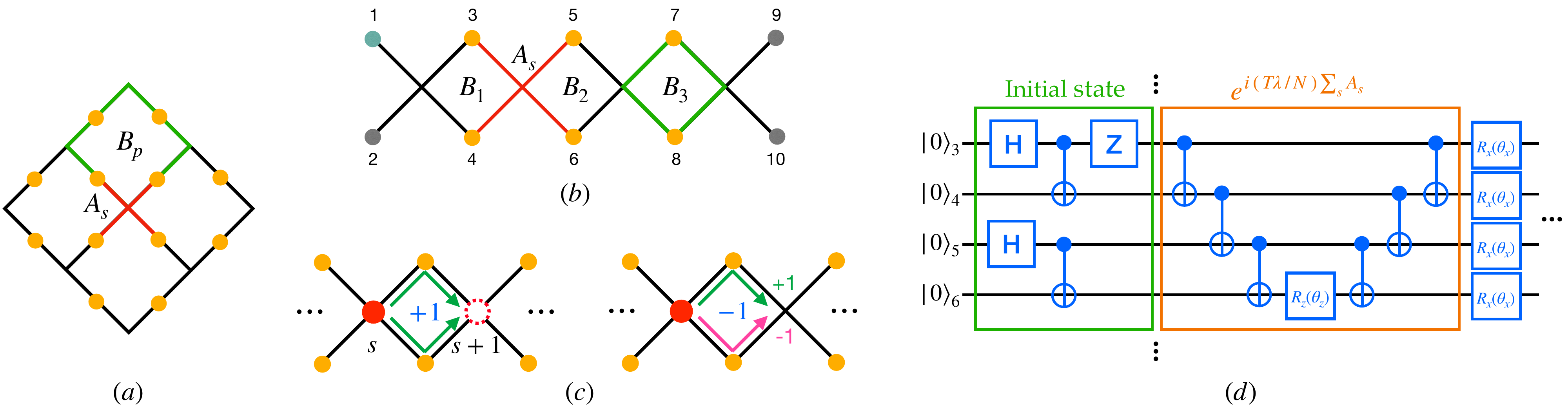}
    \caption{
    (a) Schematics of the original 2D toric code where the star operators $A_s$ live on the vertices and plaquette operator $B_p$ live on the faces of a square lattice. (b) Geometry of a toric ladder of $4$ stars $A_s$ (one highlighted in red).
    Plaquettes $B_p$ (one highlighted in green) are sandwiched between adjacent stars. 
    The four boundary spins are fixed to be three up (gray) and one down (teal).
    (c) A spinon initialized at a given star $s$ ($A_s = -1$), upon the action of a transverse field, is able to hop to an adjacent star only if there is no vison on the intervening plaquette ($B_p = +1$), due to perfect destructive interference when a vison is present ($B_p = -1$). 
    (d) Partial circuit layout to emulate a Mach-Zehnder interferometer using a toric ladder Eq.~\eqref{eq:hamiltonian}. The circuit consists of two stages: (1) preparation of an initial state with/without a vison (green box); (2) implementation of the Trotterized Hamiltonian evolution of Eq.~\eqref{eq:hamiltonian}; the execution of one time evolution step of the star operator is shown in the orange box. In all cases, the system is initialized with a single spinon on the leftmost star.}
    \label{fig:schematics}
\end{figure*}
Kitaev's toric code~\cite{Kitaev2003} consists of star,  $A_s=\prod_{i\in s}\sigma^z_i$, and plaquette, $B_p=\prod_{i\in p}\sigma^x_i$, operators acting on $S={1/ 2}$ spins on the links of a lattice as illustrated in Fig.~\ref{fig:schematics} (a). The ground state of the model is a simultaneous eigenstate of star and plaquette operators with eigenvalue $+1$ (see Eq.~\eqref{eq:hamiltonian} and related discussion). The system harbors a canonical example of topological order and anyonic excitations: stars (plaquettes) with eigenvalue $-1$ are the excitations known as spinons (visons). While these two types of excitations are (separately) bosonic, braiding one around the other imparts a phase of $\pi$ to the wave function. 
In the toric code, excitations are static, and their mutual statistics only come into play when externally manipulated. 
However, applying small external fields—small enough to preserve the energy gap and avoid transitioning to a conventional paramagnet with confined spinons and visons—can make these excitations itinerant.
\begin{figure*}[!t]
    \centering
    \includegraphics[width=0.87\linewidth]{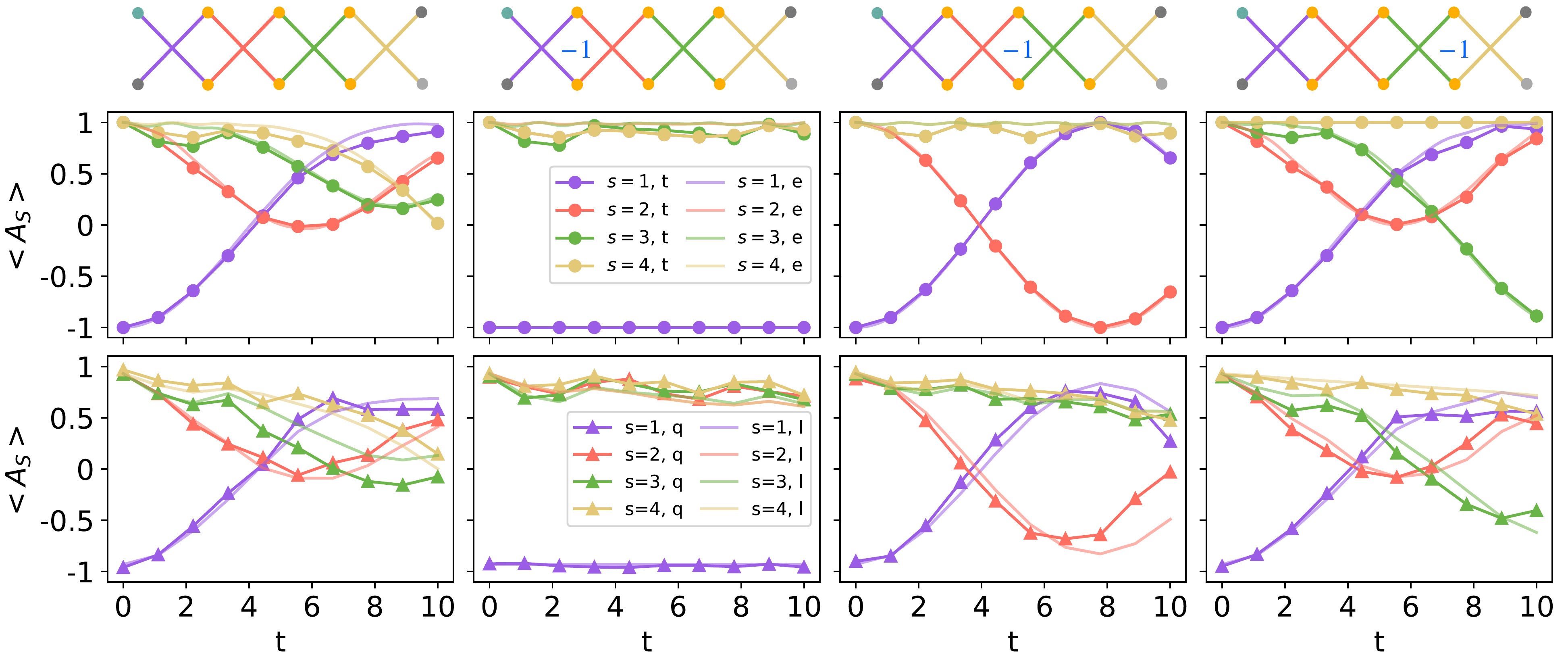}
    \caption{Spinon propagation in the background of four initial vison configurations as indicated in the schematics on the top row: (i) no vison, $B_{1,2,3} = 1$ (first column); (ii) a single vison on the left plaquette, $B_1 = -1$, $B_2 = 1$, $B_3 = 1$ (second column); (iii) a single vison on the center plaquette, $B_1 = 1$ $B_2 = -1$, $B_3 = 1$ (third column); (iv) a single vison on the right plaquette $B_1 = 1$ $B_2 = 1$, $B_3 = -1$ (last column). The expectation values of star operators $\langle A_s \rangle$ at all sites $s=1,2,3,4$ are calculated to track the location of the spinon. The upper panel shows the results obtained from exact diagonalization `e' (solid thin lines) contrasted with Trotterization `t' (solid circles with lines). The lower panel shows the IonQ Aria simulation results `q' (solid triangles with lines) compared with Lindbladian dynamics `l' (solid thin lines) with noise strength $\gamma = 0.003$. The IonQ results are obtained by measuring each qubit in the Pauli Z-basis. For a toric ladder of 4 stars, we take $500$ measurements at each time step to calculate the expectation value $\langle A_s \rangle$.}
    \label{fig:results}
\end{figure*}
%

The semionic statistics then manifests in a prominent way~\cite{PhysRevB.101.064428, Hart_2021, Zhou_2023}, underpinned by a mechanism that can be most easily understood at the single plaquette level. Consider introducing an external field that keep thes visons static and allows the spinons to propagate. When a vison occupies a plaquette of the square lattice, the coherent motion of a spinon from one corner of the plaquette to the diagonally opposite one is governed by the interference of the two trajectories along each side of the diagonal. Mutual semionic statistics causes perfect destructive interference, leading to a vanishing spinon wave function at the opposite corner. Therefore, spinon motion is entirely obstructed by the presence of the vison. In a 2D system, each trajectory will have a symmetric counterpart with respect to the diagonal, leading to destructive interference, preventing the spinon from reaching points along the diagonal line behind the vison.
Such interference blockade effects are sensitive to noise and decoherence, and challenging to observe in current digitized quantum computing devices (see Fig.~\ref{fig:2d_results} and the related discussion).

Here we focus on a minimal model, a 1D analog of the toric code, where the effects of mutual statistics on quasiparticle motion are most dramatic, and hence most easily detected. Consider the `toric ladder' illustrated in Fig.~\ref{fig:schematics}(b) for $10$ spins (which can be trivially extended to a ladder of arbitrary length). It encompasses four star operators  $A_s=\prod_{i\in s}\sigma^z_i$, and three plaquettes (here composed of $2$ spins each), sandwiched between adjacent stars. 
The four spins at the edge of the ladder, labeled $i=1,2,9,10$ in Fig.~\ref{fig:schematics} (b), are fixed to be three up and one down in the $z$ basis, to enforce an odd number of spinons (stars with eigenvalue $-1$) in the system, with the lowest energy sector containing a single spinon. 
We define the Hamiltonian 
\begin{equation}
H = - \lambda \sum_s A_s - \Gamma \sum_j \sigma_j^x 
\, , 
\label{eq:hamiltonian}
\end{equation}
where in the following $\lambda=1$ is our reference dominant energy scale, and $\Gamma \ll 1$ can be assumed positive without loss of generality. The Hamiltonian commutes with three plaquette operators, $B_p = \sigma^x_{2p+1} \sigma^x_{2p+2}$ for $p=1,2,3$ in Fig.~\ref{fig:schematics} (b), 
and $B_p = \pm 1$ identifies the presence/absence of a vison on plaquette $p$. 

Note that the star operators $A_s$ do not commute with the Hamiltonian due to the presence of a transverse field $\Gamma$. This has two effects: spinon hopping and creation/anihilation of spinon pairs. Given that the spinon energy cost is dominant in our system, 
different spinon number sectors are well separated in energy and hopping is the leading effect.
The sector of primary interest in our work will be the single spinon sector. Importantly, a spinon at a given star can hop to neighboring stars by applying the transverse field on the spin at the top or the bottom leg of the ladder, as flipping a single spin reverses the sign of $A_s$ for the two stars to which the spin belongs. The overall motion is then given by the interference of the two trajectories, on either side of the intervening plaquette $p$: if $B_p=+1$ (no vison), constructive interference allows the motion, but if $B_p=-1$ (vison), destructive interference blocks the spinon motion along the ladder, as shown in Fig.~\ref{fig:schematics} (b). 

To test this, we examine quantum quenches starting from an initial state with a single spinon localized on the leftmost star for four different vison configurations: no vision and a single vison in each of the three plaquettes.
In the limiting scenario where spinon number sectors are infinitely separated in energy, the Hamiltonian time evolution leads to a spinon delocalized across all stars in the absence of a vison, and delocalized on all stars to the left of the vison in the presence of a vison, with the stars to the right of the vison remaining inaccessible.
We will then compare this ideal Mach-Zehnder interferometer setup with scenarios where energy scales are finite and noise is present, relevant to realistic quantum devices.
%
%

We emulate the Mach-Zehner interferometer in the IonQ Aria 1 quantum device available in the AWS Braket (see device details in the Supplementary Material). Specifically, we first prepare an initial state with a spinon, with and without a vison, and then simulate the Trotterized time evolution according to the Hamiltonian in Eq.~\eqref{eq:hamiltonian} of the quantum quench discussed in the previous section. 

The starting state of the Aria quantum device $\vert 0 \rangle^{\otimes 10}$ (i.e., all qubits polarised in the positive $z$ direction) satisfies the condition of all star operators $A_s$ having eigenvalue $+1$. We then set all plaquette operators $B_p$ to have eigenvalue $+1$ by applying a Hadamard gate to each odd qubit, namely along the top leg of the ladder in Fig.~\ref{fig:schematics}(b), and following it by a sequence of CNOT gates, see Fig.~\ref{fig:schematics}(d). We manually pin the boundary qubits such that qubit-$1$ is in the down state and qubit-$2,9,10$ are all in the up state (implemented in practice by reducing the star operators $A_{s=1,4}$ to be two-qubit instead of four-qubit terms, appearing with opposite sign in the Hamiltonian). This boundary configuration initializes a spinon on the first star and constrains the total number of spinons to be odd. 
The vison can be created by applying a $Z$ gate to one of the qubits (on the top or bottom leg) at the desired plaquette. The partial circuit layout to implement the initial state is shown in the green box in Fig.~\ref{fig:schematics}(d). 

To simulate the time evolution of a given initial state according to the Hamiltonian Eq.~\eqref{eq:hamiltonian} in the digital quantum device, we use Trotterization: 
\begin{equation}
    e^{-iHt} \simeq 
\left( e^{i \frac{t \lambda}{n} \sum_s A_s} \; e^{i \frac{t\Gamma}{n} \sum_j \sigma_j^x} \right)^n
\, ,
\quad 
n \gg 1
\end{equation}
where $\hbar$ is set to $1$. This becomes exact in the limit of number of Trotter steps $n \rightarrow \infty$. However, the noisy nature of quantum devices and their limited coherence time impose a strong preference for shallow circuits. For this reason, we numerically compare the exact diagonalization time evolution with Trotterized evolution to find the smallest number of steps which gives us an approximation error less than $10\%$ (see Supplementary Material for details). We then use this as our optimal value $n_{\text{opt}}$ in the quantum device implementation. For the $10$-spin toric ladder and total time $T = 10$ (with $\lambda = 1$ and $\Gamma=0.1$), we find the optimal number of Trotter steps to be $n_{\text{opt}} = 9$. Fig.~\ref{fig:results} (top row) shows the comparison of the time evolution simulated from exact diagonalization and Trotterization. 

To implement the time evolution associated with the star operators in the Hamiltonian, $e^{i \frac{t \lambda}{n} A_s}$, we decompose it into a series of quantum gates consisting of CNOT gates and $\text{Rz}(\theta_z)$ gates that rotate the qubit about the $z$-axis by an angle $\theta_z = \frac{- t \lambda}{n}$, shown in the orange box in Fig.~\ref{fig:schematics}(d). The transverse field evolution $e^{i \frac{t\Gamma}{n} \sigma_j^x}$ is just the $\text{Rx}(\theta_x)$ gate that rotates the qubit about the $x$-axis by an angle $\theta_x = \frac{- t \Gamma}{n}$. 
%
%
\begin{figure}[!t]
    \centering
    \includegraphics[width=0.93\linewidth]{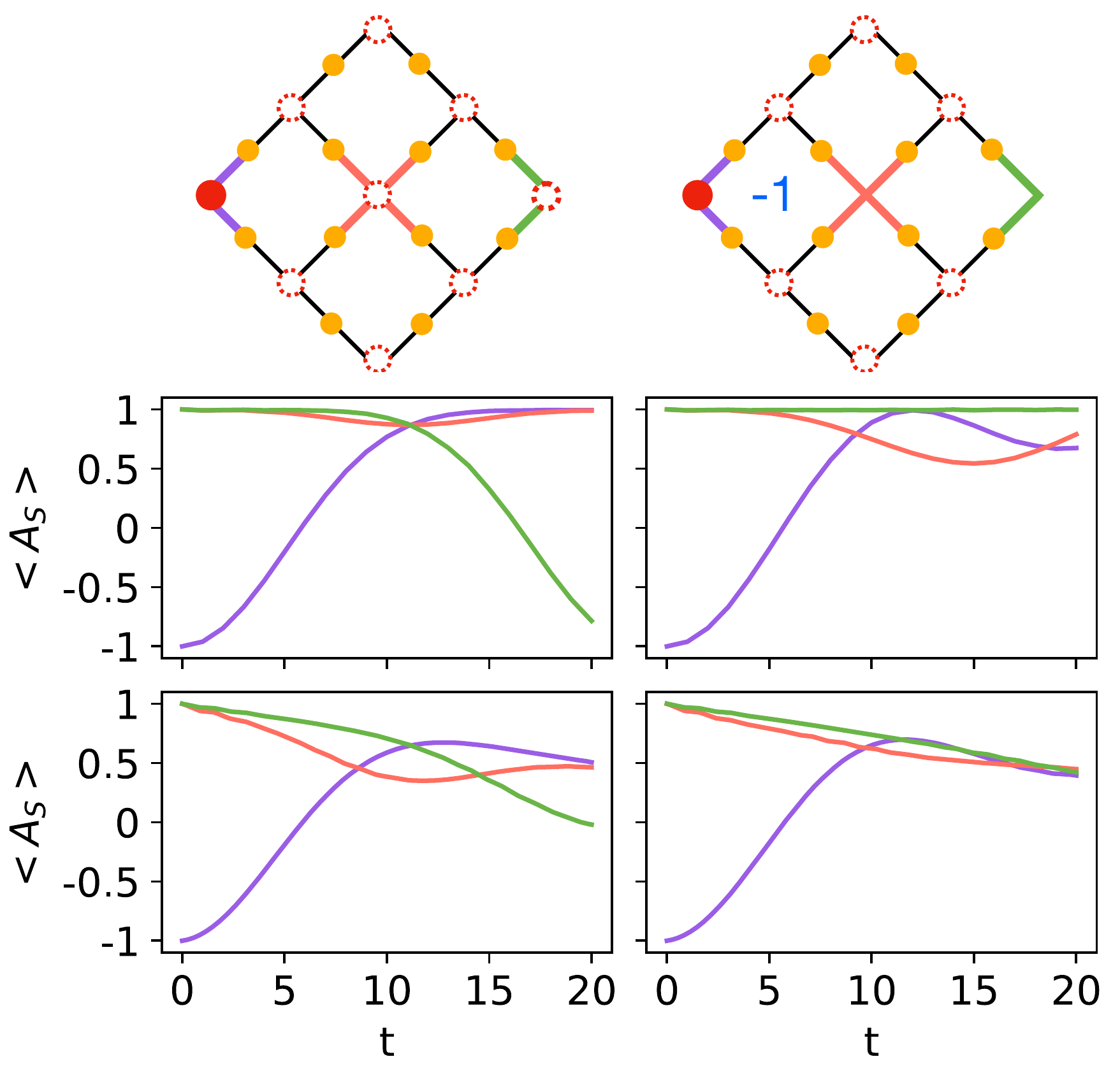}
    \caption{
    Spinon propagation in a 2D surface code of $3 \times 3$ stars ($2 \times 2$ plaquettes) with two vison configurations as shown in the schematics on the top row: i) no vison; ii) one vison at the leftmost plaquette. The boundary spins are fixed to be all up except one to be down to enforce an odd number of spinons, and are omitted in the schematics for simplicity. The expectation values of star operators $\langle A_s \rangle$ at three sites (colored in the schematics) are calculated. 
    The upper panel shows the results obtained from exact diagonalization, while the lower shows the Linbaldian dynamics with noise strength $\gamma = 0.003$.
    }
    \label{fig:2d_results}
\end{figure}

%
We show in Fig.~\ref{fig:results} the simulation results of a single spinon propagating in the background of different initial vison arrangements. The total time is $T = 10$, and at each intermediate time step, we measure the expectation value of the star operators $\langle A_s \rangle$ at all sites $s=1,2,3,4$, effectively tracking the location of the spinon as a function of time. The upper row displays the exact time evolution in comparison to the Trotterized one. We reaffirm the Mach-Zehnder interferometry-type behavior discussed earlier: the extent to which the spinon is able to propagate is determined by the interference pattern arising from the mutual semionic statistics of the spinon and the vison. The second plot shows that the spinon propagation stops in the presence of an adjacent vison, contrasted to the first plot where no vison is present, allowing the spinon to propagate freely. The third and fourth plots depict the scenarios where the vison is one or two plaquettes away, causing the spinon to oscillate between the first two or three stars, respectively. 

The lower panels in Fig.~\ref{fig:results} show the simulation results from the IonQ Aria quantum device. As time evolves, the spinon propagates to the rest of the ladder in the absence of visons (bottom first panel).
The quantum coherent oscillations of the spinon propagating across the ladder, notably visible in the absence of noise (top first panel), are mildly suppressed in the quantum device by an overall decay envelope (eventually relaxing to a maximally mixed state~\footnote{Notice in the bottom second panel from the left in Fig.~\ref{fig:results} that the value of $\langle A_1 \rangle$ does not appear to decay, contrary to the other $\langle A_{s=2,3,4} \rangle$. This is due to the compiler’s optimization cancelling CNOT gates associated with the qubit 3 and 4. A detailed explanation is provided in the Supplementary Material.}). 
In the second to fourth lower panels in Fig.~\ref{fig:results}, we observe clear signatures of the spinon blockade caused by the presence of a vison in one of the central plaquettes, suggesting that there is indeed enough coherence for the mutual semionic statistics to manifest in the system. 

In order to understand more closely the behavior observed in the quantum environment provided by IonQ, we couple our model in Eq.~\eqref{eq:hamiltonian} to an inhomogeneous isotropic single-spin bath {\it a la} Lindblad~\footnote{Note that an isotropic bath introduces dephasing of the spinon hopping amplitudes as well as decay in both the spinon and vison number. All processes are indeed needed for the simulations to agree well with the IonQ results.}: 
\begin{eqnarray}
\dot{\rho} 
= 
- i [H,\rho] 
+  \sum_i \gamma_i \left(
  \sigma^x_i \rho \sigma^x_i + \sigma^y_i \rho \sigma^y_i + \sigma^z_i \rho \sigma^z_i - 3\,\rho
\right) 
\label{eq:lindblad}
\end{eqnarray}
where the sum is over all the spins, and $\gamma_i$ is the bath coupling strength corresponding to spin $i$. The values $\gamma_i$ were chosen by optimizing the agreement with IonQ results~\footnote{To accurately reflect the quantum simulations on IonQ devices after optimized compilation, we set the noise strength of the first two spins to zero, $\gamma_3 = \gamma_4 = 0$, as a results of the CNOT cancellation mentioned in a previous footnote. See Supplementary Material for the detailed explanation of the inhomogeneous bath model and an estimate of the optimal value of $\gamma$ based on squared error.} 

Note that in the IonQ simulations we are unable to attain perfect initialization values $\langle A_1 \rangle = -1$ and $\langle A_2 \rangle = \langle A_3 \rangle = \langle A_4 \rangle = 1$. 
As a result, our initial states prepared in the quantum device could only achieve a fidelity of approximately $93\%$. To account for this imperfect initial state, in the Lindbladian simulations we use the working assumption that the initial density matrix of the system is a mixture: $p \rho_0 + (1-p) \mathbb{I}/N$, with $p\approx0.93$; here $\rho_0$ is the pure density matrix of the desired initial state, and $\mathbb{I}/N$ is the density matrix of a totally mixed state. 


While our choice of bath is not drawn from any realistic attempt to describe accurately the source of noise in IonQ quantum devices, the agreement that we observe throughout the range of initial conditions and time scales probed in our work is remarkable. With this better understanding of the device noise, we discuss the possibility of observing the interferometric signatures of fractional exchange statistics directly in the 2D surface code. We show in Fig.~\ref{fig:2d_results} the $3\times3$ lattice geometry and the results for spinon propagation in the absence and presence of a vison, using exact numerical and noisy Linbladian dynamics. Similar to the ladder case, the boundary spins are fixed to allow us to work in the sector with approximately a single spinon, with the dynamics generated by a time-independent toric-code Hamiltonian with a small transverse field. The noisy simulations use the noise parameters extracted from matching the IonQ results with the Linbladian model for the toric ladder.

While similar to the ladder case, the spinon blockade effect in present in the exact simulation, unlike the ladder results presented in Fig.~\ref{fig:results}, it is clear from Fig.~\ref{fig:2d_results} that the same device noise level washes out the difference between spinon propagation with and without vison.

In conclusion, we developed a synthetic Mach-Zehnder interferometer, suitable for current noisy quantum devices, to probe the fractional exchange statistics of excitations in topological states. Specifically, we implemented the toric ladder, a quasi-1D version of the toric code, which displays striking interference effects due to mutual semionic statistics. This results in a spinon blockade effect driven by vison excitations, and we show that this phenomenon is highly dependent on quantum coherence, making it challenging to observe unambiguously in 2D.

Notwithstanding the noisy quantum environment, our setup successfully revealed interferometric signatures of semionic statistics -- an effect directly tied to quantum coherence -- on the IonQ Aria 1 quantum processor. The Mach-Zehnder interferometer offers key advantages for detecting fractional statistics: the statistical signatures arise naturally from the system’s nonadiabatic quantum evolution under a time-independent Hamiltonian, eliminating the need for complex control schemes for adiabatic braiding. Moreover, the exchange statistics can be extracted without auxiliary qubits, unlike the approach adopted, e.g, in Ref.~\onlinecite{Satzinger2021}. 

Beyond serving as an alternative synthetic platform for investigating anyonic statistics relevant to spin liquids, our findings offer insight into quantum computing devices' coherence length and timescales. This goal is achieved by directly examining the decay rate of spinon blockade, a phenomenon that would persist indefinitely in a noise-free system. Comparing our quantum hardware results with numerical calculations, we have validated the accuracy of a Lindbladian noise model in capturing the noise characteristics of the IonQ device.
%
%

{\it Acknowledgement} - 
We are grateful to Orazio Scarlatella for many useful discussion and for generously sharing with us his expertise on open quantum systems and Lindblad evolution. A.R. thanks Boston University's Physics Department for hospitality. This work was funded in part by the Engineering and Physical Sciences Research Council (EPSRC) grants No.~EP/T028580/1 and No.~EP/V062654/1 (C.C.), by DOE Grant No.~DE-FG02-06ER46316 (C.Ch.), and by NSF Grant No. DMR-1945395 (A.R.). Research at Perimeter Institute is supported in part by the Government of Canada through the Department of Innovation, Science and Industry Canada and by the Province of Ontario through the Ministry of Colleges and Universities. 
%
%

\bibliographystyle{apsrev4-2}
\bibliography{reference.bib}

\end{document}


\title{Supplementary Material}
\maketitle
\onecolumngrid
\section{IonQ quantum device layout\label{app:ionq}}
The IonQ quantum devices are ion-trap-based digital quantum machines, where qubits are realized using ionized ytterbium atoms, offering all-to-all connectivity~\cite{ionq}. Our experiment was conducted on the IonQ Aria device, accessed via AWS Braket (specifically Aria 1). The Aria device features 25 qubits and supports most common quantum gates, which can be mapped to its native gate set. Its average gate fidelities are approximately 99.04\% for 1-qubit gates and 98.87\% for 2-qubit gates, with gate durations of about 135 µs and $600\,\mu$s, respectively. The device's T1 (energy relaxation) time ranges from 10 to 100 seconds, while the T2 (dephasing) time is around 1 second~\cite{ionq_aria}.

\section{Optimal Trotterization
\label{app:trotterization}
}
Here we elaborate on the derivation of the optimal number of Trotter steps, $n_{\rm opt}$, used in the quantum device implementation. The selection of this optimal number takes into account two considerations: 1) the Trotter approximation error tends to $0$ in the limit of large number of Trotter steps; 2) the noisy nature of quantum devices gives a preference for shallow circuits, and therefore small number of Trotter steps, to reduce loss of coherence / thermalization. In an attempt to balance these two conflicting trends, we look for an optimal number of Trotter steps, $n_{\rm opt} = 9$ for a time evolution up to $T=10$ (see below), in order to achieve sufficient accuracy ($\lesssim 10\%$) with contained accumulated noise. 

\begin{figure}[!h]
    \includegraphics[width=0.45\columnwidth]{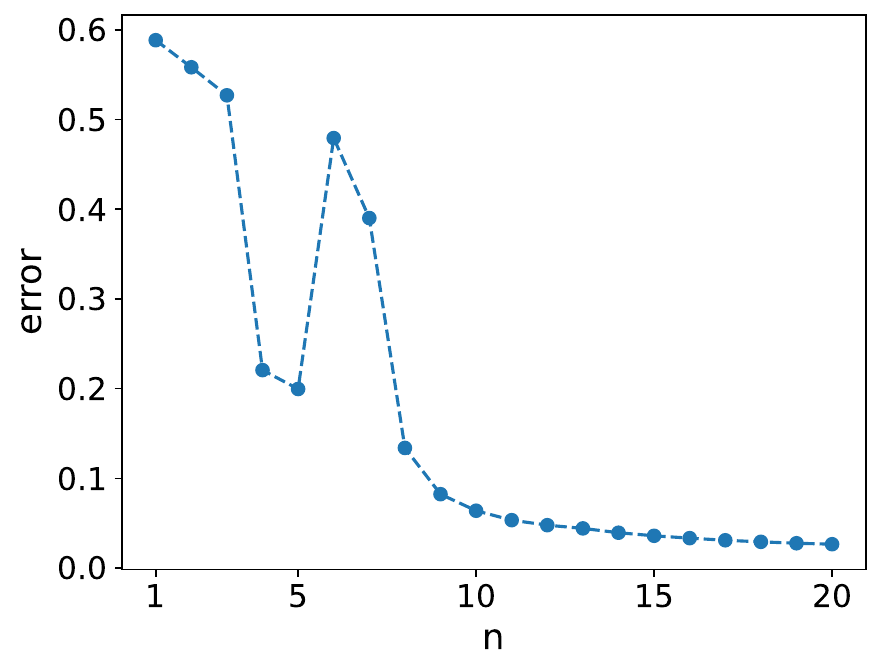}
    \caption{The Trotter approximation error versus the number of Trotter steps $n$. The error decrease with significant fluctuations down to about $0.08$ at $n=9$, and smoothly decays toward $0$ as $n$ keeps increasing.}
    \label{fig:n_opt}
\end{figure}

In Fig.~\ref{fig:n_opt}, we show the Trotter approximation error as a function of the number of Trotter steps $n$. The error is calculated by taking the absolute value of the difference between $A_s^{({\rm e})}(t)$ evaluated from exact diagonalization and $A_s^{({\rm t})}(t)$ from Trotterization, averaged over all four stars $s=1,2,3,4$ and the four vison configurations discussed in the main text, with the spinon initialized at $s=1$. As expected, the Trotter error gradually decays to $0$ as $n$ increases. We choose $10\%$ as a working error threshold and we see that this is achieved for $n \geq 9$, which then sets our optimal number of Trotterisation steps at $n_{\rm opt} = 9$. 
%
%

\section{Lindblad evolution
\label{app:lindblad}
}
\begin{figure}[h]
    \centering
    \includegraphics[width=0.9\linewidth]{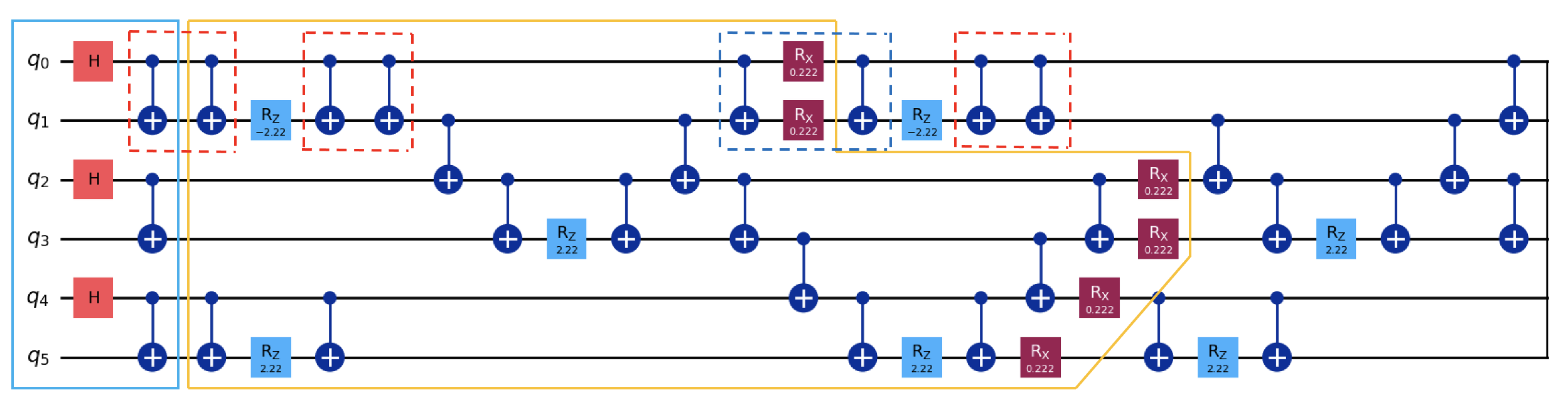}
    \caption{Circuit diagram: the blue solid box marks the initial state preparation (for the case of no visons), the orange solid box encircles one trotter step, the red dashed boxes show the exact CNOT cancellations, and the blue dashed box represents a near-cancellation of CNOT gates.}
    \label{fig:circuit_cancellations}
\end{figure}

\begin{figure}[h]
    \centering
    \includegraphics[width=0.45\linewidth]{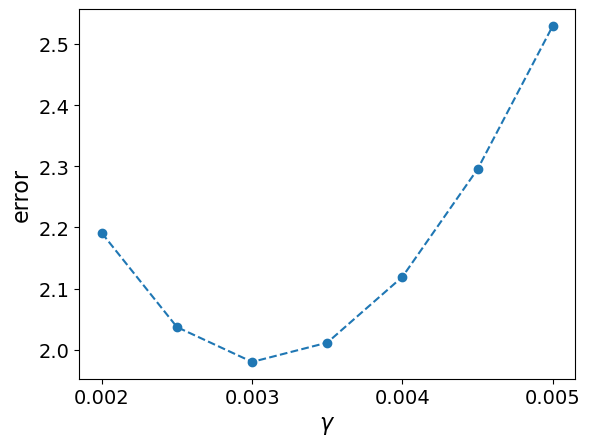}
    \caption{Square error from Lindblad simulation and IonQ hardware data summed over all stars and vison configurations.}
    \label{fig:optimal_gamma}
\end{figure}

We effectively model the noise in the quantum device IonQ by introducing independent Markovian heat baths to each spin, described by the Lindblad Master equation. Under the assumption of isotropic noise in spin space, the corresponding Lindblad equation, Eq.~(2) in the main text, is given by
\begin{equation}
\dot{\rho} 
= 
- i [H,\rho] 
+ \sum_i \gamma_i \left(
  \sigma^x_i \rho \sigma^x_i + \sigma^y_i \rho \sigma^y_i + \sigma^z_i \rho \sigma^z_i - 3\,\rho
\right) 
\, , 
\end{equation}
where the sum is over all the spins, and $\gamma_i$ is the coupling strength to the bath for spin i. In circuit language, this introduces a single-qubit depolarization channel for each qubit, and the strength of depolarization primarily depends on the number of two-qubit gates involving that qubit. From the circuit shown in Fig.~\ref{fig:circuit_cancellations}, it is clear that there are many consecutive CNOT pairs (red dash boxes) on $q_0$ and $q_1$ canceling out in optimised compilation. Another recurrent pattern on $q_0$ and $q_1$ is marked with the blue dash box, where two small rotation gates Rx are sandwiched by two CNOT gates. This produces a near-cancellation since the rotation angles are small and the whole block is close enough to the identity. In contrast, any blocks involving Rz gates cannot be regarded as a near-cancellation, as the rotation angle of Rz is ten times higher that of Rx. Therefore, it should be clear from the circuit that all CNOTs on $q_0$ and $q_1$ (except the last one in the last Trotter step) either cancel or nearly-cancel each other, leading to negligible depolarization compared to the other four qubits, where the cancellations are mostly spoiled by the extra insertion of CNOT control or target gates. Motivated by this observation, we introduce zero dephasing for $q_0$ and $q_1$, and a homogeneous Lindblad dephasing strength $\gamma$ for $q_2$, $q_3$, $q_4$ and $q_5$. The optimal value of $\gamma$ is then obtained by computing and minimizing the square error between $\langle A_s \rangle$ from the Lindblad simulation data and the IonQ hardware data, summing over all stars and all vison configurations. The error, shown in Fig.~\ref{fig:optimal_gamma}, indicates that the optimal value is $\gamma = 0.0030 \pm 0.0005$. 

Our last remark is on the implementation method of our Lindblad model. The main difficulty in simulating Eq.~(3) lies in the presence of both operators acting to the left and to the right of $\rho$, preventing any factorisation of the right hand side that could transform it into a matrix eigenvalue problem. To overcome this, we introduce the vectorised form of the density matrix $\vec{\rho}$, by stacking the rows of its matrix representation. The Lindblad equation is then recast as:
%
\begin{equation}
\dot{\Vec{\rho}} = 
\left\{ 
-i (H \otimes I - I \otimes H^T) 
+ \sum_i \gamma_i \left[
  \sigma^x_i \otimes (\sigma^x_i)^T 
  + \sigma^y_i \otimes (\sigma^y_i)^T 
  + \sigma^z_i \otimes (\sigma^z_i)^T 
  - 3 I \otimes I
\right] 
\right\} \vec{\rho} 
= \mathcal{L} \vec{\rho} 
\, , 
\end{equation}
where $\otimes$ denotes the standard tensor product, and $\mathcal{L}$ is sometimes referred to as the Lindbladian superoperator. Now that the Master equation is put into 
this new form, 
the vectorised density matrix ${\vec{\rho}}$ can be trivially evolved in the eigenbasis of $\mathcal{L}$ and can then be reshaped into the conventional matrix form $\rho$. In practice, to capture the trotterization steps as in the circuit implementation, we first apply the trotterized Hamiltonian and then apply the inhomogeneous Lindblad dephasing part with an amplitude $\gamma dt$ at the end of each trotter step. The diagonalization follows the same vectorization protocol illustrated above.

\bibliographystyle{apsrev4-2}
\bibliography{reference.bib}